\documentclass[letterpaper,english,reprint, aps,prl]{revtex4-1}
\usepackage[T1]{fontenc}
\usepackage[latin9]{inputenc}
\usepackage{color}
\usepackage{amsmath}
\usepackage{amssymb}
\usepackage{graphicx}

\makeatletter

\pdfpageheight\paperheight
\pdfpagewidth\paperwidth

\providecommand{\tabularnewline}{\\}

\usepackage{tikz}
\usetikzlibrary{arrows}
\usetikzlibrary{shadows}
\tikzset{
  every overlay node/.style={
    draw=white,anchor=north west,
  },
}
\usetikzlibrary{positioning, decorations.pathmorphing}

\renewcommand{\fnum@figure}{FIG.~\thefigure}

\makeatother

\usepackage{babel}
\begin{document}
\preprint{APS/123-QED}
\title{Independent dimensional phase transition on a two-dimensional Kuramoto
model with matrix coupling}
\author{Chongzhi Wang}
\author{Haibin Shao}
\author{Dewei Li}
\email{dwli@sjtu.edu.cn}

\affiliation{Department of Automation, Shanghai Jiao Tong University, Key Laboratory
of System Control and Information Processing, Ministry of Education
of China, Shanghai Engineering Research Center of Intelligent Control
and Management, Shanghai, 200240, China}
\date{\today}
\begin{abstract}
The high-dimensional generalization of the one-dimensional Kuramoto
paradigm has been an essential step in bringing about a more faithful
depiction of the dynamics of real-world systems. Despite the multi-dimensional
nature of the oscillators in these generalized models, the interacting
schemes so far have been dominated by a scalar factor unanimously
between any pair of oscillators that leads eventually to synchronization
on all dimensions. As a natural extension of the scalar coupling befitting
for the one-dimensional case, we take a tentative step in studying
numerically and theoretically the coupling mechanism of $2\times2$
real matrices on two-dimensional Kuramoto oscillators. One of the
features stemmed from this new mechanism is that the matrix coupling
enables the two dimensions of the oscillators to separate their transitions
to either synchronization or desynchronization which has not been
seen in other high-dimensional generalizations. Under various matrix
configurations, the synchronization and desynchronization of the two
dimensions combine into four qualitatively distinct modes of position
and motion of the system. We demonstrate that as one matrix is morphed
into another in a specific manner, the system mode also switches correspondingly
either through continuous or explosive transitions of the order parameters,
thus mimicking a range of behaviors in information science and biology.
\end{abstract}

\maketitle

Autonomous behavior of large ensembles of interacting entities has
been an observation captured both in the realms of nature and of artificial
creations \citep{winfree1967biological,wiener2019cybernetics}. Proposed
as a mathematically-tractable model to address this fact and probably
as the nontrivial many-body problem in its simplest form \citep{watanabe1994constants},
the Kuramoto paradigm was thoroughly studied for the exemplary phenomena
it induces on the all-to-all sinusoidally coupled one-dimensional
oscillators on the unit circle \citep{kuramoto2003chemical,strogatz2000kuramoto,van1993lyapunov}.
Among other things, the synchronization phase transition hints at
similar processes in many problems of physical or engineering background
such as the Josephson junction arrays \citep{watanabe1994constants},
the XY model with quenched randomness \citep{uezu2015correspondence},
and those of biological background \citep{buck1968mechanism,acebron2005kuramoto}.
Due to this versatility in lower dimensions, the classic Kuramoto
model was rapidly gaining generalizations in many ways possible, for
which there is in particular, the high-dimensional Kuramoto model
where the phase variables are replaced by unit vectors that are coupled
proportionally to the sine of the vectors' displacement angle \citep{zhu2013synchronization,chandra2019continuous};
or in a multi-layer setting, information of an oscillator spreads
into different layers, and inter-layer as well as intra-layer interactions
take place \citep{zhang2015explosive}. Regardless, a single scalar
variable spans the parameter space that tunes the coupling strength
for all dimensions where each eventually secures a synchronization,
hence no discernable long-term behaviors are displayed thereon as
along as the coupling is sufficiently strong.
\begin{figure}
\usetikzlibrary {shapes.geometric,fadings}
\begin{tikzpicture}
    \definecolor{osc1}{RGB}{72,61,139}
    \definecolor{osc2}{RGB}{220,220,220}
    \definecolor{txt}{RGB}{245,245,220}
    \definecolor{Beige}{RGB}{220,220,220}
    \definecolor{Surface}{RGB}{177,162,159}
    \definecolor{link}{RGB}{77,57,0}
    \definecolor{ring}{RGB}{255,255,240}

   \path [fill=Beige!50] (2.5,3.1) to (3.5,4.6) to (9.5,4.6) to (8.5,3.1) to (2.5,3.1);
   \path [fill=Beige!50] (2.5,5) to (3.5,6.5) to (9.5,6.5) to (8.5,5) to (2.5,5);

    \node (n21) at (4,5.8) [circle,shading=radial,inner color=osc1,outer color=osc2,draw=Surface!20,minimum size=6mm,node font=\sffamily,text=txt] {2};
    \node (n22) at (4,4) [circle,shading=radial,inner color=osc1,outer color=osc2,draw=Surface!20,minimum size=6mm,node font=\sffamily,text=txt] {2};
    \node (n11) at (5.9,5.5) [circle,shading=radial,inner color=osc1,outer color=osc2,draw=ring,very thick,minimum size=6mm,node font=\sffamily,text=txt] {1};
    \node (n12) at (5.9,3.7) [circle,shading=radial,inner  color=osc1,outer color=osc2,draw=Surface!20,minimum size=6mm,node font=\sffamily,text=txt] {1};
    \node (n31) at (7.9,6) [circle,shading=radial,inner  color=osc1,outer color=osc2,draw=Surface!20,minimum size=6mm,node font=\sffamily,text=txt] {3};
    \node (n32) at (7.9,4.2) [circle,shading=radial,inner  color=osc1,outer color=osc2,draw=Surface!20,minimum size=6mm,node font=\sffamily,text=txt] {3};

    \draw [-,dashed,ultra thick,color=link!40] (n21) -- (4,5.1);
    \draw [-,dashed,ultra thick,color=link] (4,5) -- (n22);
    \draw [-,dashed,ultra thick,color=link!40] (n11) -- (5.9,5);
    \draw [-,dashed,ultra thick,color=link] (5.9,5) -- (n12);
    \draw [-,dashed,ultra thick,color=link!40] (n31) -- (7.9,5);
    \draw [-,dashed,ultra thick,color=link] (7.9,5) -- (n32);
    \draw [-,ultra thick,color=link] (n21) -- (n11);
    \node (p11) at (5,5.9) {$p_{11}$};
    \draw [-,ultra thick,color=link] (n31) -- (n11);
    \node (q11) at (6.9,5.95) {$q_{11}$};
    \draw [-,ultra thick,color=link] (n22) -- (n12);
    \node (p12) at (5,4.1) {$p_{12}$};
    \draw [-,ultra thick,color=link] (n32) -- (n12);
    \node (q12) at (6.9,4.15) {$q_{12}$};

    \node (dim1) at (9.4,5.8) {$\theta_{i1}$ dimension};
    \node (dim2) at (9.4,4) {$\theta_{i2}$ dimension};


\end{tikzpicture}

\caption{\emph{The matrix coupling mechanism}. Illustration of how the three
oscillators and their dimensions are interacting, given that $\theta_{1}$
is connected with $\theta_{2}$ and $\theta_{3}$ weighed by matrices
$P=\begin{bmatrix}p_{11} & p_{12}\protect\\
p_{21} & p_{22}
\end{bmatrix}$ and $Q=\begin{bmatrix}q_{11} & q_{12}\protect\\
q_{21} & q_{22}
\end{bmatrix}$ respectively. Solid lines connect variables that are differenced
before being acted on by the sine function, while dashed lines connect
two dimensions of the same oscillator that are not explicitly interacting.
By examining the dynamics of $\theta_{11}$, one finds that the terms
$\sin(\theta_{j1}-\theta_{11})$ and $\sin(\theta_{j2}-\theta_{12}),j=2,3$
are weighed by the first row of the coupling matrices $P,Q$, namely,
$p_{11},p_{12},q_{11},$ and $q_{12}$, then summed to determine frequency
$\dot{\theta}_{11}$. The second row of the matrices, in a similar
manner, would be contributing to the dynamics on the $\theta_{i2}$
dimension.}

\label{fig:FIG1}
\end{figure}
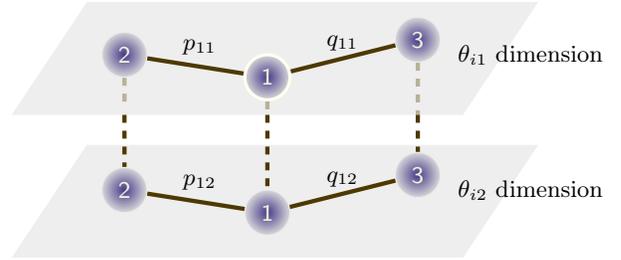

If we revisit the generalization process and take the simplest possible
multi-dimensional case to work with, for a three dimensional unit
vector, the oscillator is left with two degrees of freedom that are
fully described by the two angle variables in the sphere coordinates.
In this Letter, we propose a novel interacting mechanism, the matrix
coupling, that directly operates on the differences of the angle variables
for the oscillators on the unit sphere, and report phase transitions
as well as other systematical behaviors that distinguish from other
high-dimensional Kuramoto generalizations. The proposed dynamics reads

\begin{equation}
\dot{\theta}_{i}=\left[\begin{array}{c}
\dot{\theta}_{i1}\\
\dot{\theta}_{i2}
\end{array}\right]=\left[\begin{array}{c}
\omega_{i1}\\
\omega_{i2}
\end{array}\right]+\frac{1}{N}\cdot A\sum_{j=1}^{N}{\bf sin}(\theta_{j}-\theta_{i})\label{eq:1}
\end{equation}
where both the dimensions $\theta_{i1},\theta_{i2}$ are assumed to
be angular variables of oscillator $i$, with $\omega_{i1}$ and $\omega_{i2}$
drawn from the distributions $g_{1}(\omega_{1})$ and $g_{2}(\omega_{2})$
being their natural frequencies of oscillation in that direction.
Now, instead of the averaged scalar factor, we consider an averaged,
all-to-all coupling with the matrix $A=\begin{bmatrix}a_{11} & a_{12}\\
a_{21} & a_{22}
\end{bmatrix}\in\mathbb{R}^{2\times2}$, that acts on the sum of vectors $[\sin(\theta_{j1}-\theta_{i1})\;\sin(\theta_{j2}-\theta_{i2})]^{T}$.
The matrix coupling arises naturally from the characterization of
the inter-dimensional communication amongst multi-dimensional entities,
and has been considered in the context of opinion dynamics on interdependent
topics \citep{friedkin2016network}, synchronization on coupled arrays
of LC oscillators or pendulums \citep{tuna2019synchronization} and
more \citep{TRINH2018415,zhao2016localizability}. FIG. \ref{fig:FIG1}
illustrates the mechanism of the two-dimensional Kuramoto model with
the matrix-coupling; one should notice that the dynamics on one dimension
of a specific oscillator is taking direct influences from both dimensions
of the neighboring oscillators, except the dimensions are weighed
differently by the row elements. This interaction is further clarified
if we define the complex order parameters for the two dimensions as
$\rho_{1}=\frac{1}{N}\sum_{j=1}^{N}e^{i\theta_{j1}}=\sigma_{1}e^{i\Psi_{1}},\rho_{2}=\frac{1}{N}\sum_{j=1}^{N}e^{i\theta_{j2}}=\sigma_{2}e^{i\Psi_{2}}$,
where $0\leq\sigma_{1,2}\leq1$ and $\Psi_{1,2}$ denote the average
phases; the equations of motion for $\theta_{i1},\theta_{i2}$ are
then
\begin{figure}
\includegraphics[width=8.6cm]{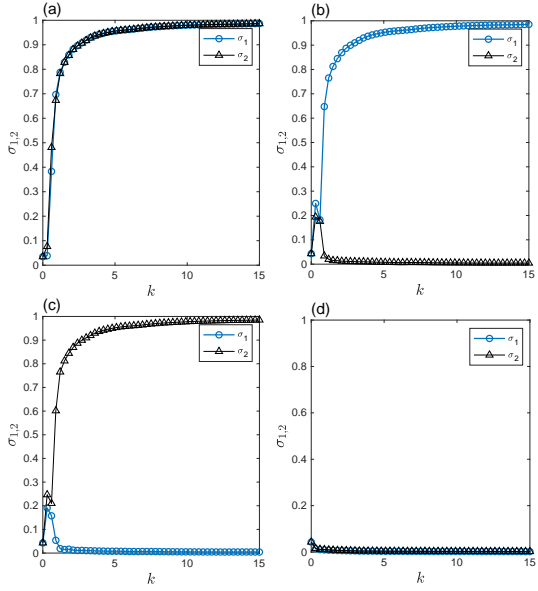}

\caption{\emph{Transitions of $\sigma_{1,2}$ under candidate matrices $A_{i}$.}
Panels (a), (b), (c), and (d) are results of direct simulation of
eqn. (\ref{eq:1}) on $N=1000$ oscillators with $A_{i}=A_{1},A_{2},A_{3},A_{4}$
respectively; each panel demonstrates the values of the order parameter
$\sigma_{1}$ (blue circle) and $\sigma_{2}$ (black triangle) during
the transition under the corresponding candidate matrix, for which
$\sigma_{1}\rightarrow1(\sigma_{2}\rightarrow1)$ indicates a high
level of synchrony in dimension $\theta_{i1}(\theta_{i2})$ while
$\sigma_{1}\rightarrow0(\sigma_{2}\rightarrow0)$ indicates desynchrony,
i.e., a uniform distribution in dimension $\theta_{i1}(\theta_{i2})$.}

\label{fig:FIG2}
\end{figure}

\begin{equation}
\begin{array}{c}
\dot{\theta}_{i1}=\omega_{i1}-a_{11}\sigma_{1}\sin(\theta_{i1}-\Psi_{1})-a_{12}\sigma_{2}\sin(\theta_{i2}-\Psi_{2}),\\
\dot{\theta}_{i2}=\omega_{i2}-a_{21}\sigma_{1}\sin(\theta_{i1}-\Psi_{1})-a_{22}\sigma_{2}\sin(\theta_{i2}-\Psi_{2}).
\end{array}\label{eq:2}
\end{equation}

\noindent We see that the instantaneous frequencies are modulated
by the weighed $\theta_{i1}$ mean-field and $\theta_{i2}$ mean-field.
It is then of question what the variation of the elements is going
to bring to the state of the system, measured by the order parameters
$\rho_{1},\rho_{2}$.

\textcolor{black}{To get down to the essentials}, we simulate on $N=1000$
oscillators whose initial conditions are identically $\theta_{i}(0)={\bf 0}$,
assuming the natural frequencies $\omega_{i1}=\omega_{i2}$ to be
drawn from the Lorentzian distribution $g_{1}(\omega)=g_{2}(\omega)=g(\omega)=\frac{1}{\pi}\frac{\gamma}{(\omega-\Omega)^{2}+\gamma^{2}}$
with symmetry center $\Omega=0$ and spread $\gamma=1$. 

\begin{figure}
\begin{tabular}{|c|c|}
\hline 
\includegraphics[width=4cm]{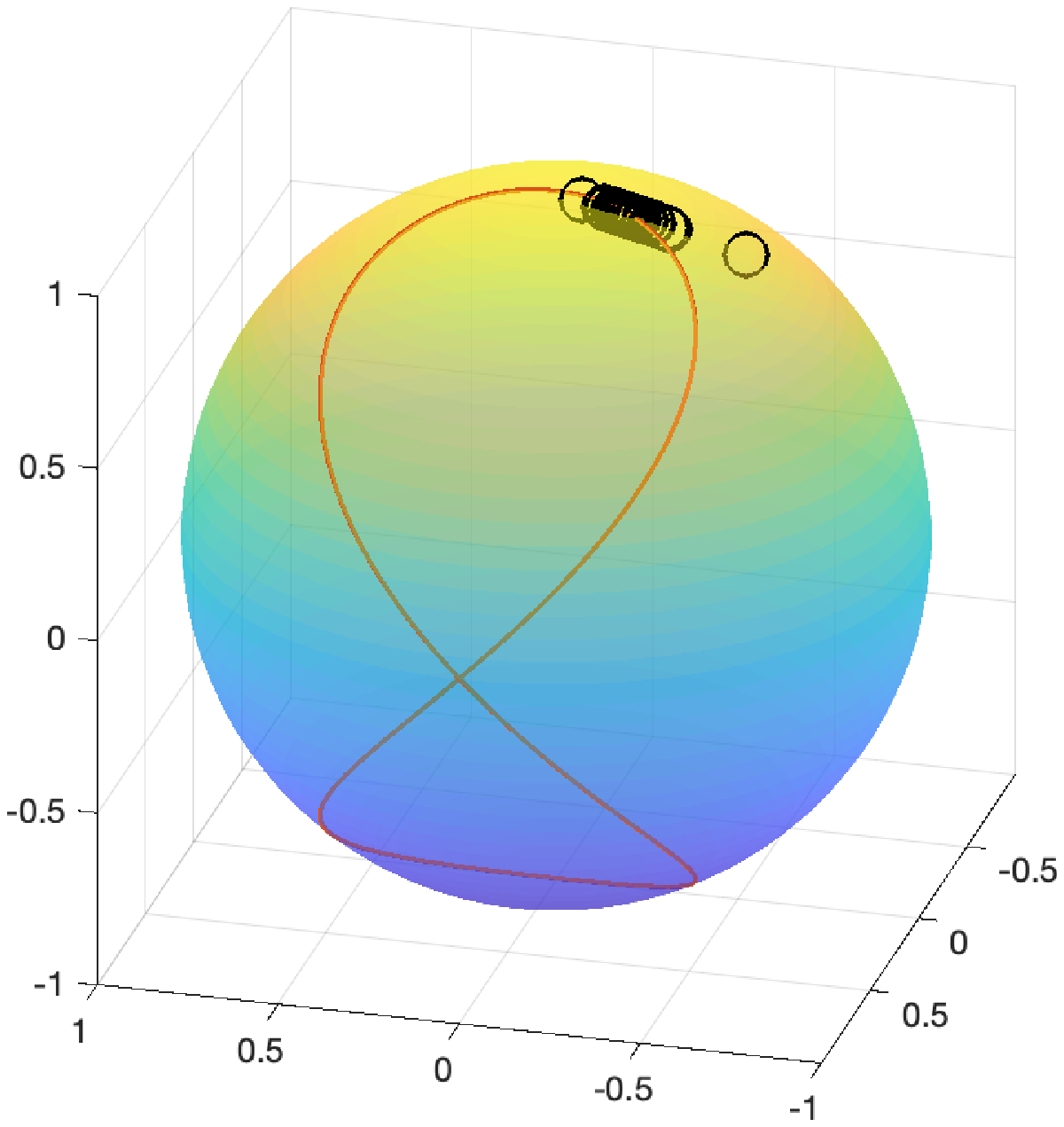} & \includegraphics[width=4cm]{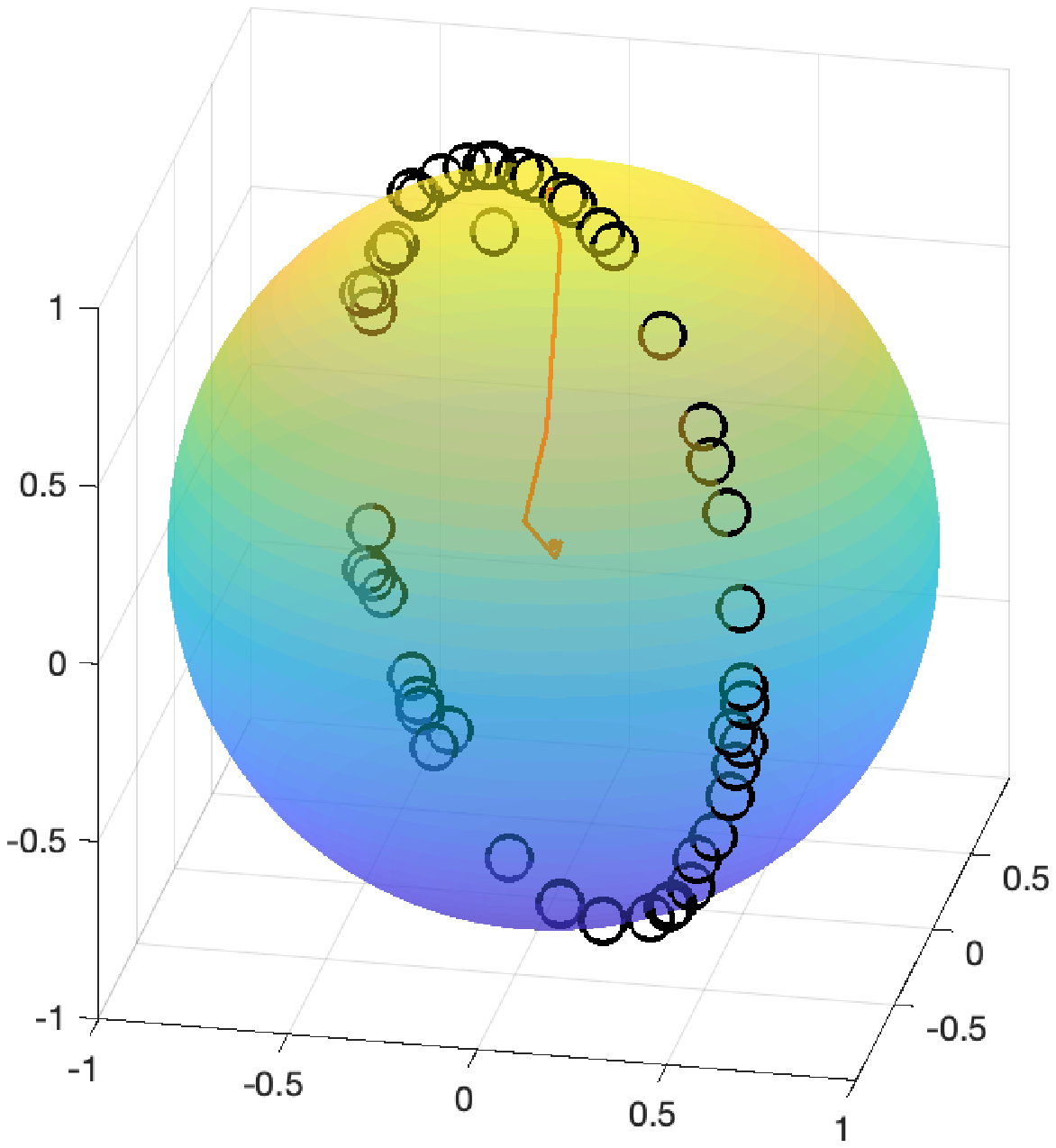}\tabularnewline
\hline 
\includegraphics[width=4cm]{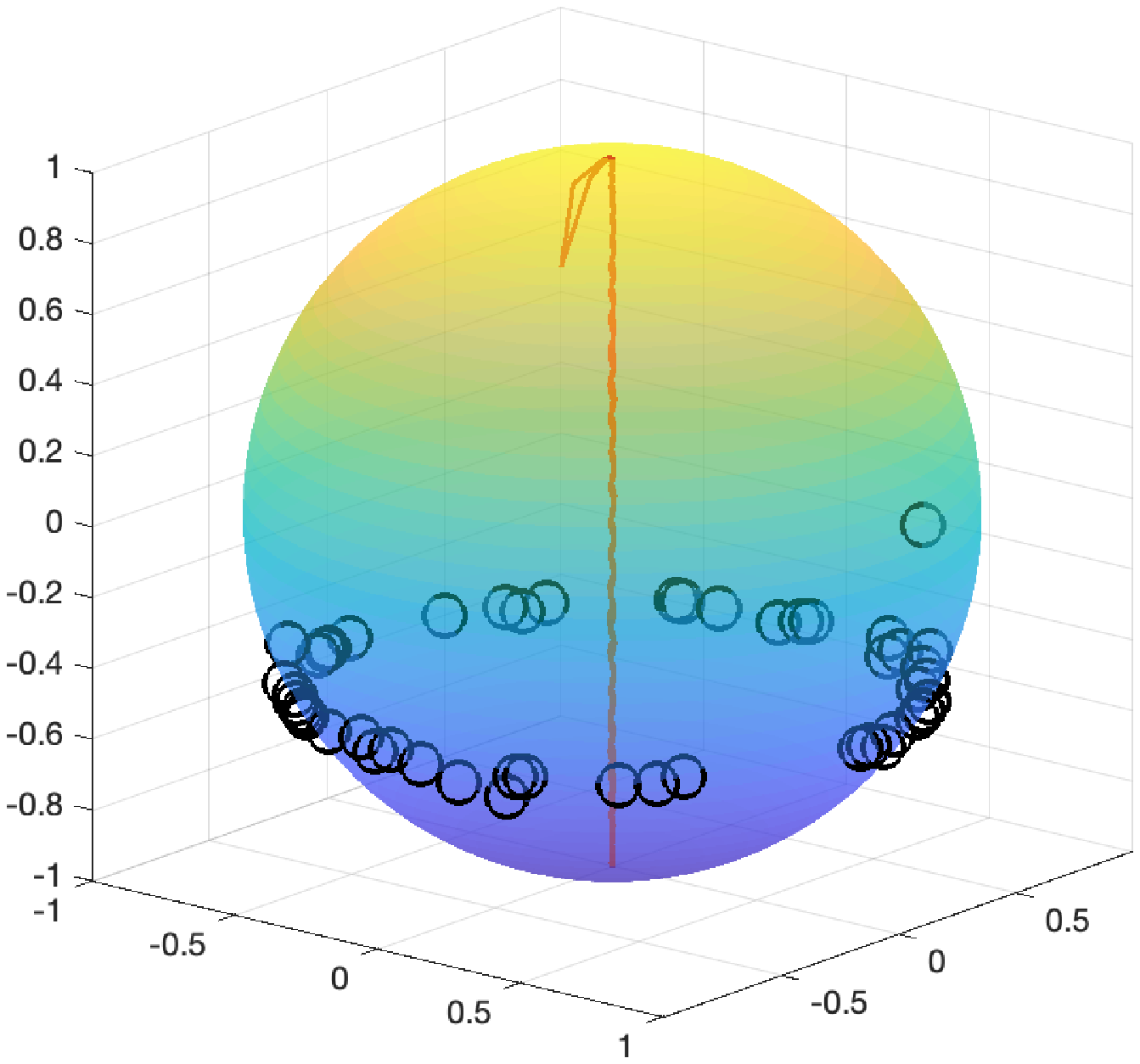} & \includegraphics[width=4cm]{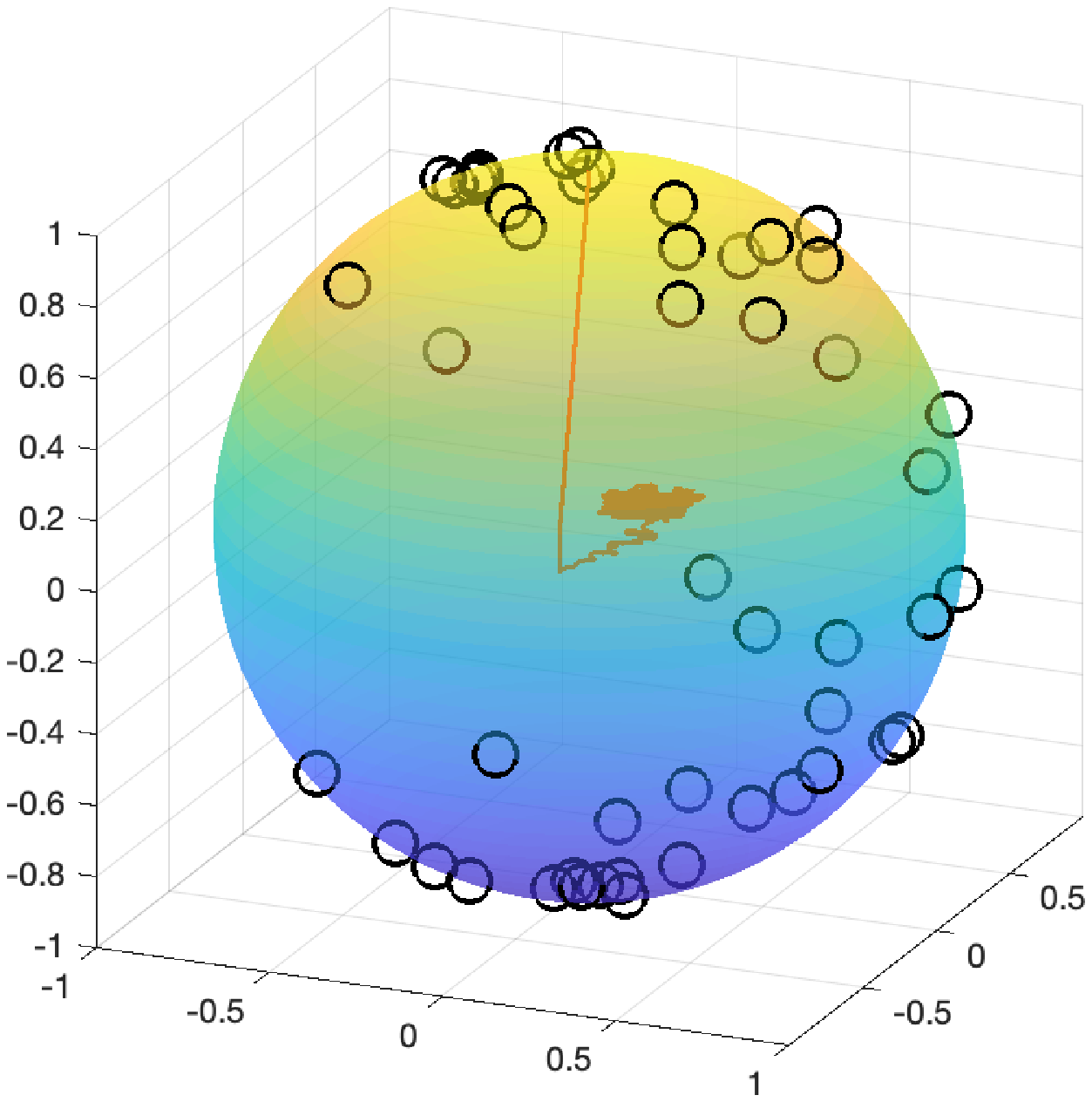}\tabularnewline
\hline 
\end{tabular}

\caption{\emph{Visualization on the unit sphere.} Take $\theta_{i1}$ as the
azimuthal angle and $\theta_{i2}$ as the polar angle, the population
of oscillators (black circles) are projected onto the surface of a
unit sphere, where the orange line stands for the trajectory of the
average position of the population w.r.t time. We illustrate with
$N=50$ oscillators whose initial conditions are identically $\theta_{i}(0)={\bf 0}.$
The four panels correspond to the four modes the system eventually
settled in \textcolor{black}{with the candidate matrices $A_{1},A_{2},A_{3},A_{4}$,}
as $k$ is tuned to $50$. The upper-left panel shows when the oscillators
are fully synchronized (in frequency) on both $\theta_{i1}$ and $\theta_{i2}$
dimensions; the oscillators remain relatively static and travel along
the orange orbit at a constant speed. Upper-right panel shows when
the oscillators are only $\theta_{i1}$ synchronized and stay disarranged
on the ring that rotates uniformly about the $z$-axis, their average
position soon approaching the origin. In the lower-left panel where
the population is only $\theta_{i2}$ synchronized, the majority distributes
on a ring that is shrinking and stretching as it moves up and down
in the $z$-direction. While for the lower-right panel, the $\theta_{i1},\theta_{i2}$
desynchronized oscillators drift across the surface and do not form
any groups or show any particular pattern. We mention, in the first
three cases, the velocity of the mean-field is visibly non-zero because
of the limitation of $N$, which renders the center of the natural
frequencies to be non-zero.}

\label{fig:FIG3}
\end{figure}

The first discovery could be exemplified by four candidate matrices
$A_{1},A_{2},A_{3},A_{4}$ we used to generate FIG. \ref{fig:FIG2},
where $A_{1}=\begin{bmatrix}4.27 & 0.11\\
7.42 & 3.20
\end{bmatrix},A_{2}=\begin{bmatrix}3.80 & 7.54\\
7.57 & 2.86
\end{bmatrix},A_{3}=\begin{bmatrix}2.86 & 7.54\\
7.57 & 3.80
\end{bmatrix},A_{4}=\begin{bmatrix}-24.11 & 6.28\\
6.27 & -17.89
\end{bmatrix}$. For the experiment, let $A=kA_{i},i\in\{1,2,3,4\}$, we adiabatically
increase $k$ from $0$ to $15$ with increment $\Delta k=0.3$ and
calculate the average of $\rho_{1,2}$ during a set period of time
as the stationary order parameters of the corresponding $k$.\textcolor{red}{{}
}We note that this scaling is fundamentally different from that in
other generalized models, as there are now four independent variables
undergoing changes in the parameter space of the system. FIG. \ref{fig:FIG2}
reports our result that for $A_{1}$ and $A_{4}$, both the $\theta_{i1}$
dimension and the $\theta_{i2}$ dimension go through qualitatively
similar transitions either from being largely desynchronized to a
complete frequency synchronization ($\sigma_{1,2}\rightarrow1$),
or to a complete desynchronization ($\sigma_{1,2}\rightarrow0$).
Yet for $A_{2}$ and $A_{3}$ whose diagonal elements are interchanged,
the transitions on $\theta_{i1}$ dimension and $\theta_{i2}$ dimension
go to opposite directions, i.e., $\sigma_{1}\rightarrow1,\sigma_{2}\rightarrow0$
or $\sigma_{1}\rightarrow0,\sigma_{2}\rightarrow1$ as the elements
of $A$ are simultaneously and sufficiently increased.

One thing to extract from these experiments is that, the matrix coupling
enables the multi-dimensional Kuramoto oscillators to separate the
transition to synchronization/desynchronization in different dimensions,
which has not been seen in other high-dimensional generalizations
\citep{chandra2019continuous,zhang2015explosive}. Depicting the combinations
of $\sigma_{1,2}$ on the unit sphere, FIG. \ref{fig:FIG3} shows
how the system sets into four qualitatively different modes of distribution
and motion with the configurations of $A_{i}$. As to how the proposed
model is able to achieve this, we mention, which confirms the results
by many other choices of $A_{i}$, that the positivity/negativity
of the two real eigenvalues $\lambda_{1,2}$ of $A$ is reflected
on $\sigma_{1,2}$ for $k>0$ sufficiently large. Demonstrating this
with the candidate matrices, for $A_{1}$ with $\lambda_{1,2}\geq0$,
the order parameters has $\sigma_{1,2}\rightarrow1$, while for $A_{4}$
with $\lambda_{1,2}\leq0$ there is $\sigma_{1,2}\rightarrow0$ (note
that only one of $\lambda_{1},\lambda_{2}$ is permitted to be zero);
meanwhile for $A_{2}$ and $A_{3}$ that satisfy $\lambda_{1}\cdot\lambda_{2}<0$,
one obtains $\sigma_{1}\rightarrow1,\sigma_{2}\rightarrow0$ or $\sigma_{1}\rightarrow0,\sigma_{2}\rightarrow1$.
\textcolor{black}{When the elements of the coupling matrix are altered
in this manner, the transition in dimension $\theta_{i1}$ or $\theta_{i2}$
experiences several minor jumps according to $\sigma_{1,2}$ even
for populations as large as $N=1000$, which has to do with the contribution
of the drifting oscillators to the order parameters. }But in all,
the transition is comparable to that of the second-order. With the
disposition of the positivity/negativity of the eigenvalues of the
coupling matrix, the system has the tendency to set into the four
modes with complete synchronization/desynchronization of the two dimensions.
This does not apply, however, when the coupling matrix $A_{i}$ \textcolor{black}{satisfies}
$a_{11}=a_{22}>0,a_{12}=a_{21}$, as we have also discovered that
for this kind of weight matrix, the positively increased $k$ only
generates transition of $\sigma_{1}\rightarrow1,\sigma_{2}\rightarrow1$.
Another intricacy arises when the eigenvalues $\lambda_{1,2}$ are
a conjugate pair, which according to our experiments, distinguishes
between $|Re(\lambda)|>1$ and $|Re(\lambda)|\leq1$. While the case
of $|Re(\lambda)|>1$ much resembles that where the eigenvalues are
real, which means $\sigma_{1,2}\rightarrow1$ as $Re(\lambda_{1,2})>1$
and $\sigma_{1,2}\rightarrow0$ as $Re(\lambda_{1,2})<-1$, the case
of $|Re(\lambda)|\leq1$ gets more eccentric and always ends up with
$\sigma_{1,2}$ settling into steady values significantly between
0 and 1. For the remaining of this work, we avert our attention from
these ramifications and focus on the coupling matrices that are well-behaved
with real eigenvalues that lead to a full synchronization/desynchronization
on the two dimensions.

Our second discovery emerges from the four modes exhibited that lead
to the inevitable question of if the system is actually capable of
switching between one mode to another by modulating the elements of
$A$, as there are obviously many ways the coupling matrix $A$ could
alter to produce dynamics apart from being steadily scaled by a factor
$k$. What we then do is to encode the system modes into binary digits
$00,01,10,11$ encouraged by the combination of $\sigma_{1,2}$ and
find representative matrices $M_{1},M_{2},M_{3},M_{4}$ for these
modes, which means $kM_{i}$ let the system set into mode $00/01/10/11$
with $k$ positive and sufficiently large, and with a uniform initial
condition $\theta_{i}(0)={\bf 0}$.

\begin{figure}
\usetikzlibrary {arrows.meta}
\begin{tikzpicture}
    \node (00) at (1,4) {00};
    \node (01) at (4,4) {01};
    \node (10) at (4,1) {10};
    \node (11) at (1,1) {11};
    \draw [arrows = {-Stealth[left]}] (00) -- (01);
	\draw [arrows = {-Stealth[left]}] (01) -- (00);
    \draw [arrows = {-Stealth[left]}] (01) -- (10);
	\draw [arrows = {-Stealth[left]}] (10) -- (01);
    \draw [arrows = {-Stealth[left]}] (10) -- (11);
	\draw [arrows = {-Stealth[left]}] (11) -- (10);
    \draw [arrows = {-Stealth[left]}] (11) -- (00);
	\draw [arrows = {-Stealth[left]}] (00) -- (11);
    \draw [arrows = {-Stealth[left]}] (00) -- (10);
	\draw [arrows = {-Stealth[left]}] (10) -- (00);
    \draw [arrows = {-Stealth[left]}] (11) -- (01);
	\draw [arrows = {-Stealth[left]}] (01) -- (11);
	\node (a) at (2.5,4.2) {(a)};\node (b) at (2.5,3.8) {(b)};
	\node (c) at (4.2,2.5) {(c)};\node (d) at (3.8,2.5) {(d)};
	\node (f) at (2.5,1.2) {(f)};\node (e) at (2.5,0.8) {(e)};
	\node (g) at (0.8,2.5) {(g)};\node (h) at (1.2,2.5) {(h)};
	\node (i) at (3.4,1.9) {(i)};\node (j) at (1.6,3.1) {(j)};
	\node (k) at (3.1,3.4) {(k)};\node (l) at (1.9,1.6) {(l)};
\end{tikzpicture}

\begin{tabular}{ccc}
\includegraphics[width=2.6cm]{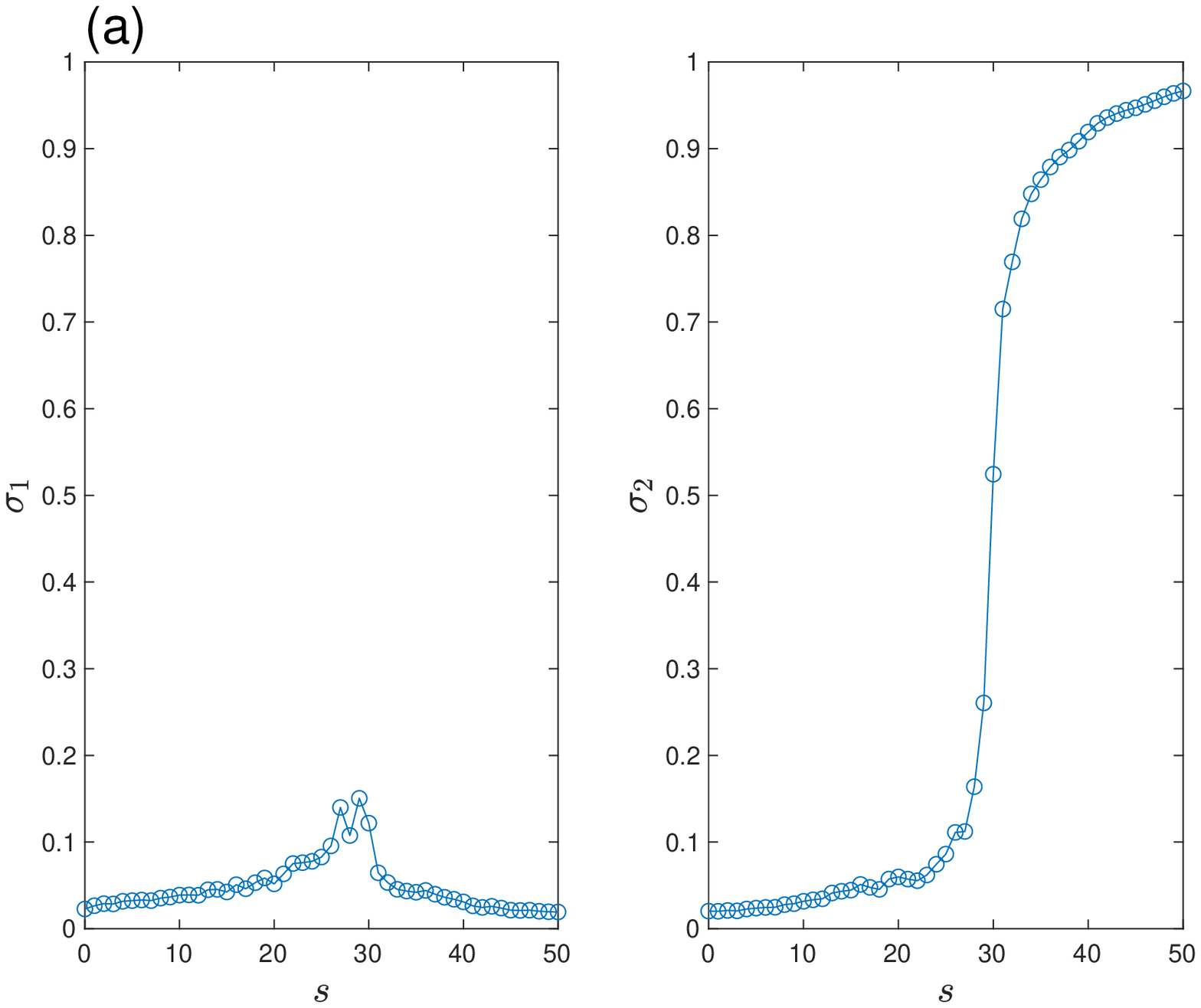} & \includegraphics[width=2.6cm]{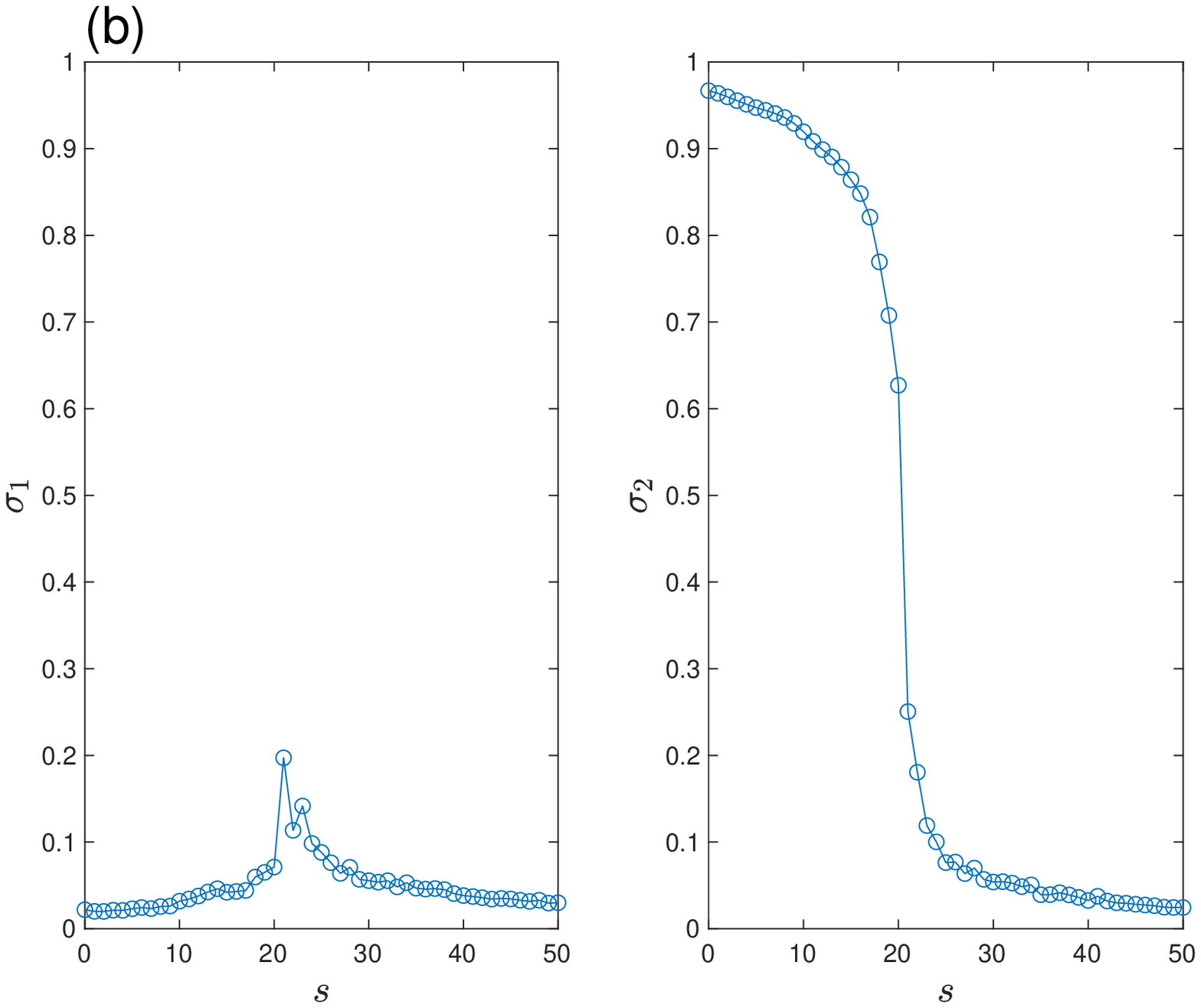} & \includegraphics[width=2.6cm]{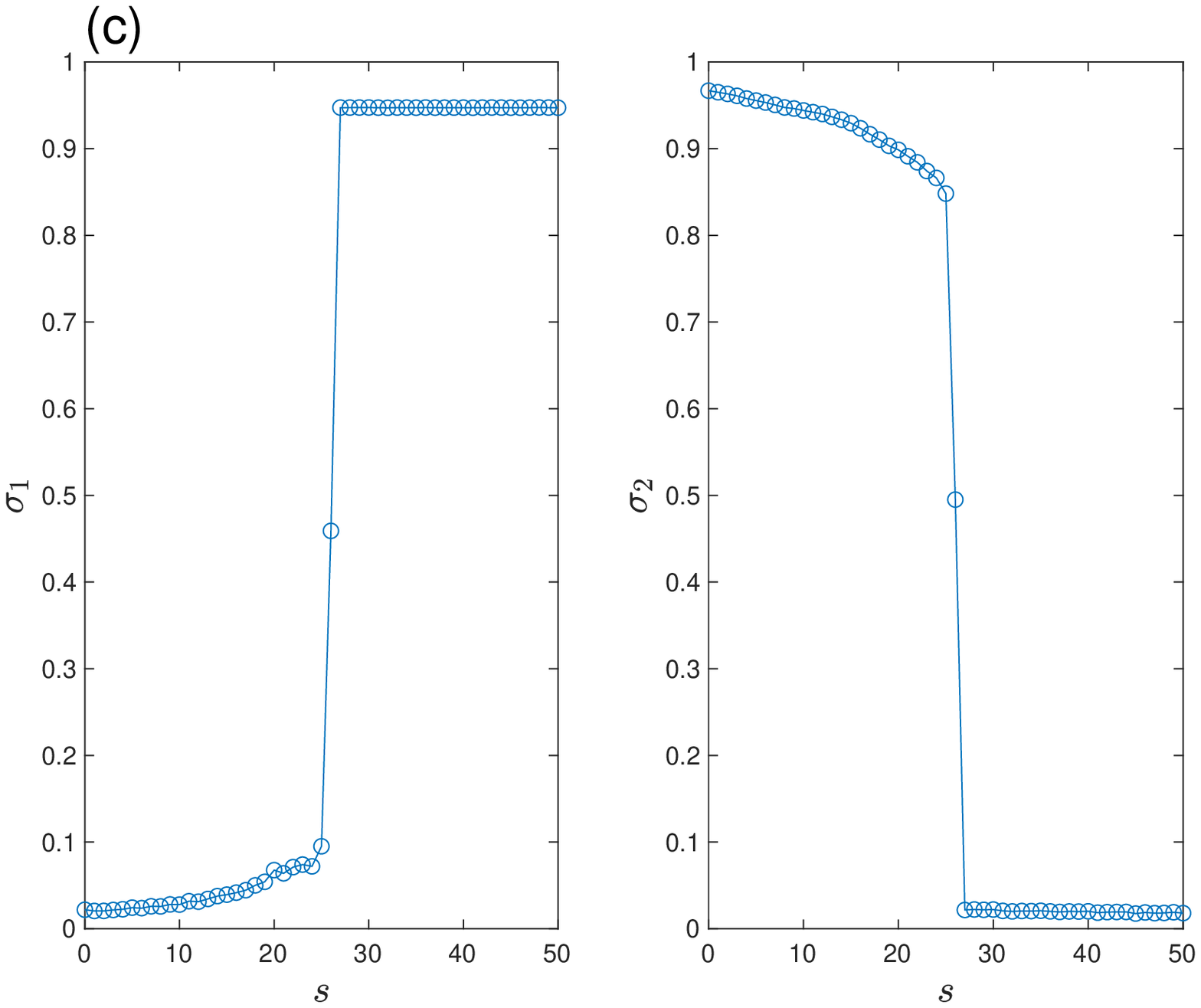}\tabularnewline
\includegraphics[width=2.6cm]{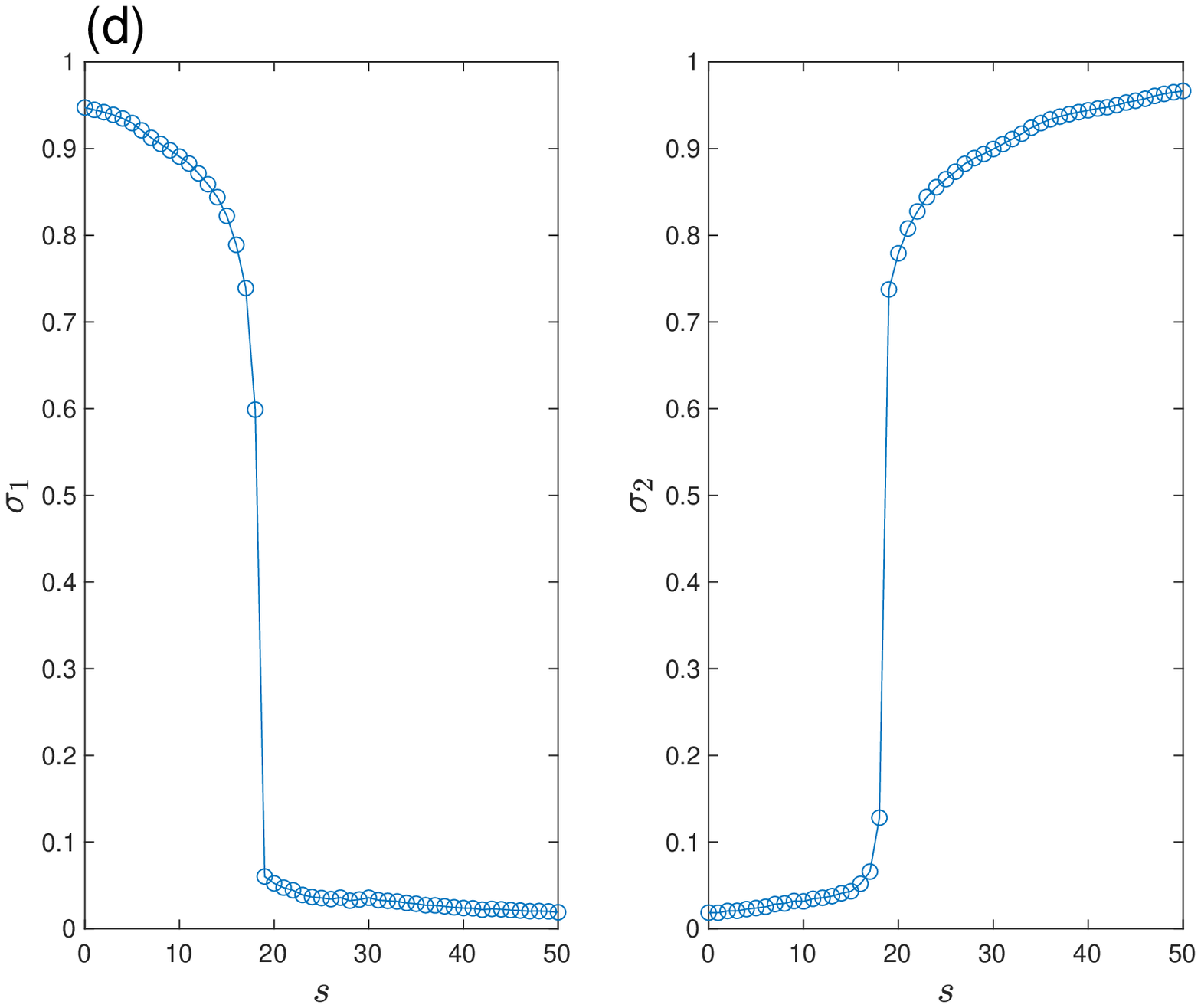} & \includegraphics[width=2.6cm]{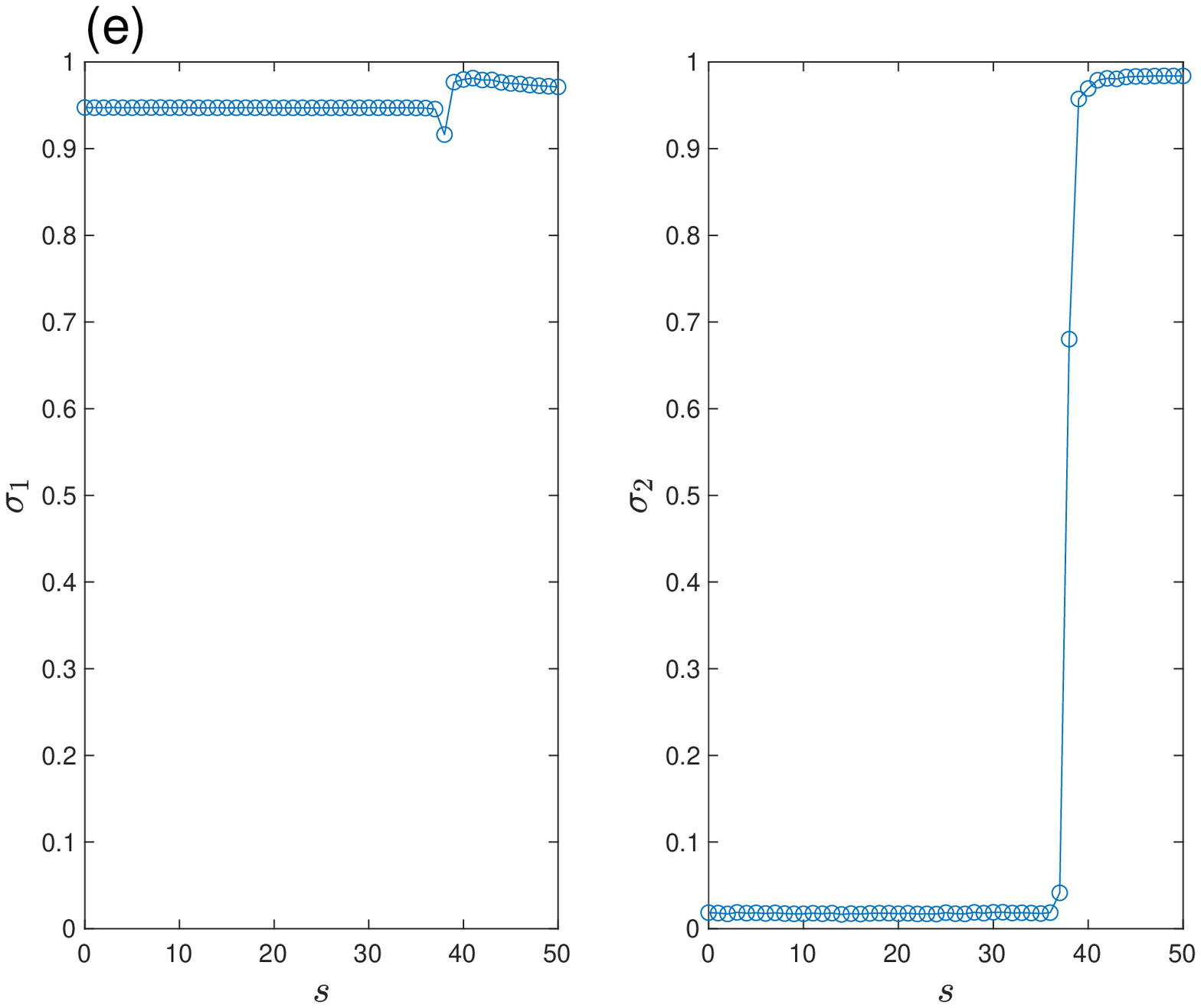} & \includegraphics[width=2.6cm]{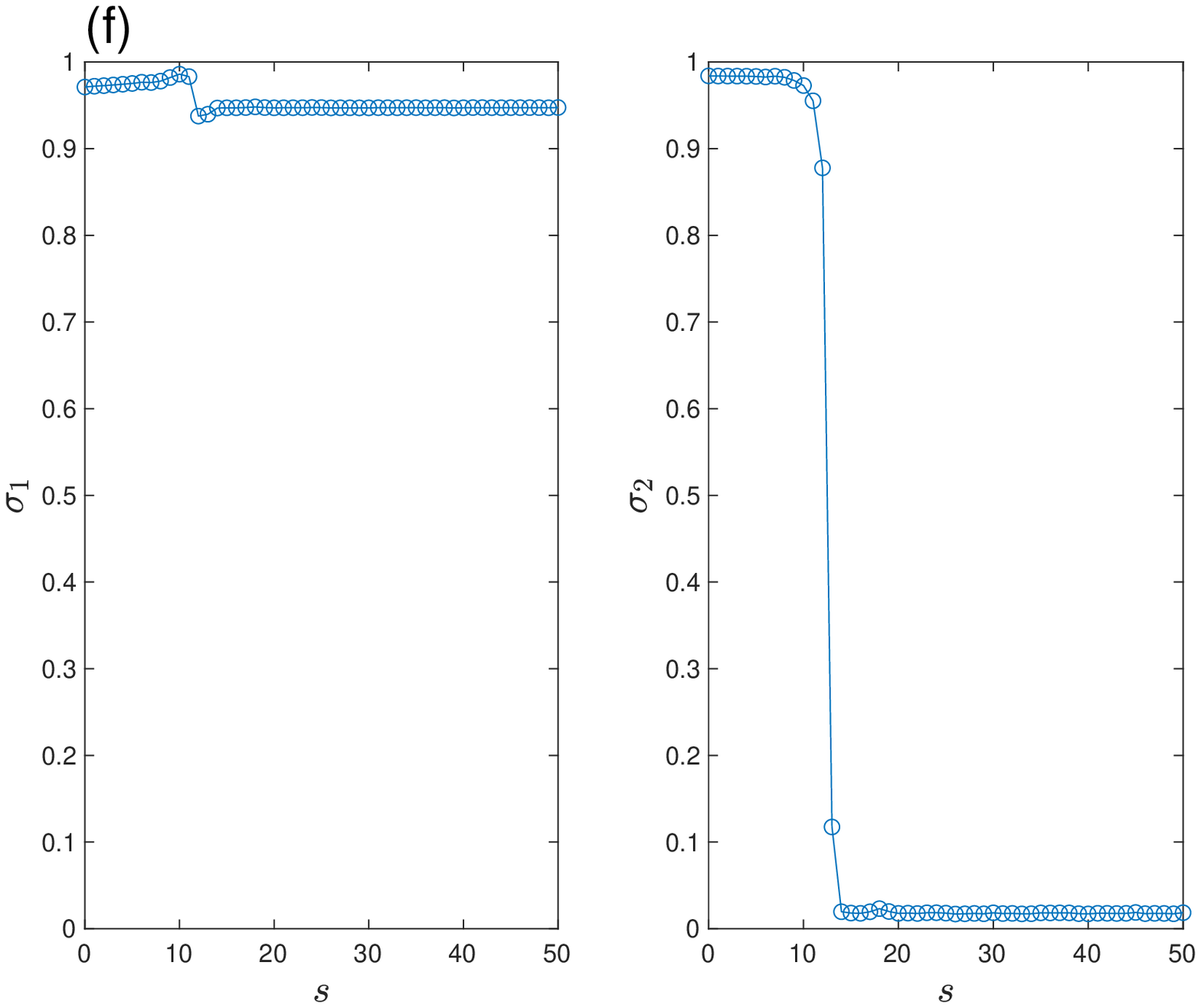}\tabularnewline
\includegraphics[width=2.6cm]{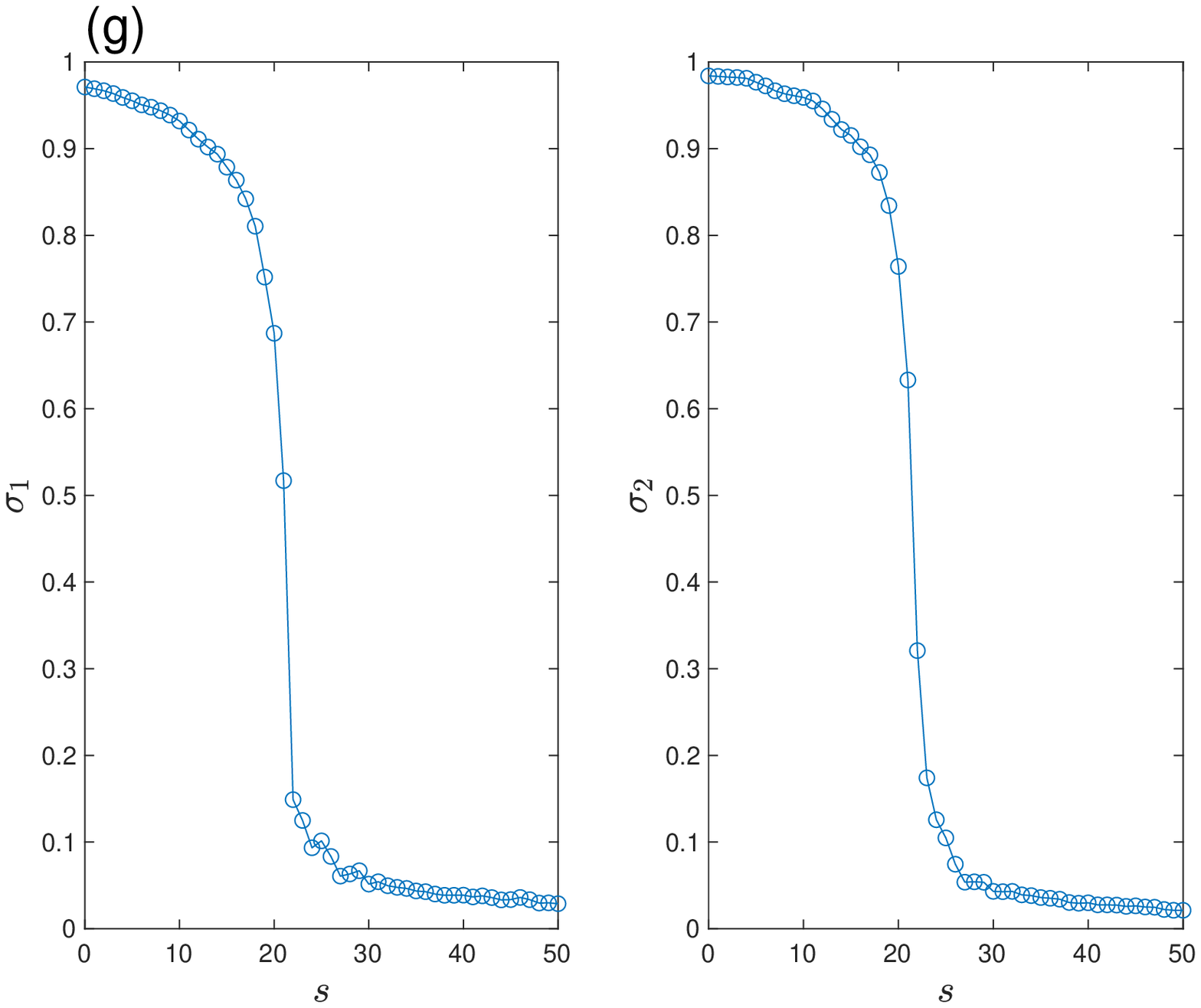} & \includegraphics[width=2.6cm]{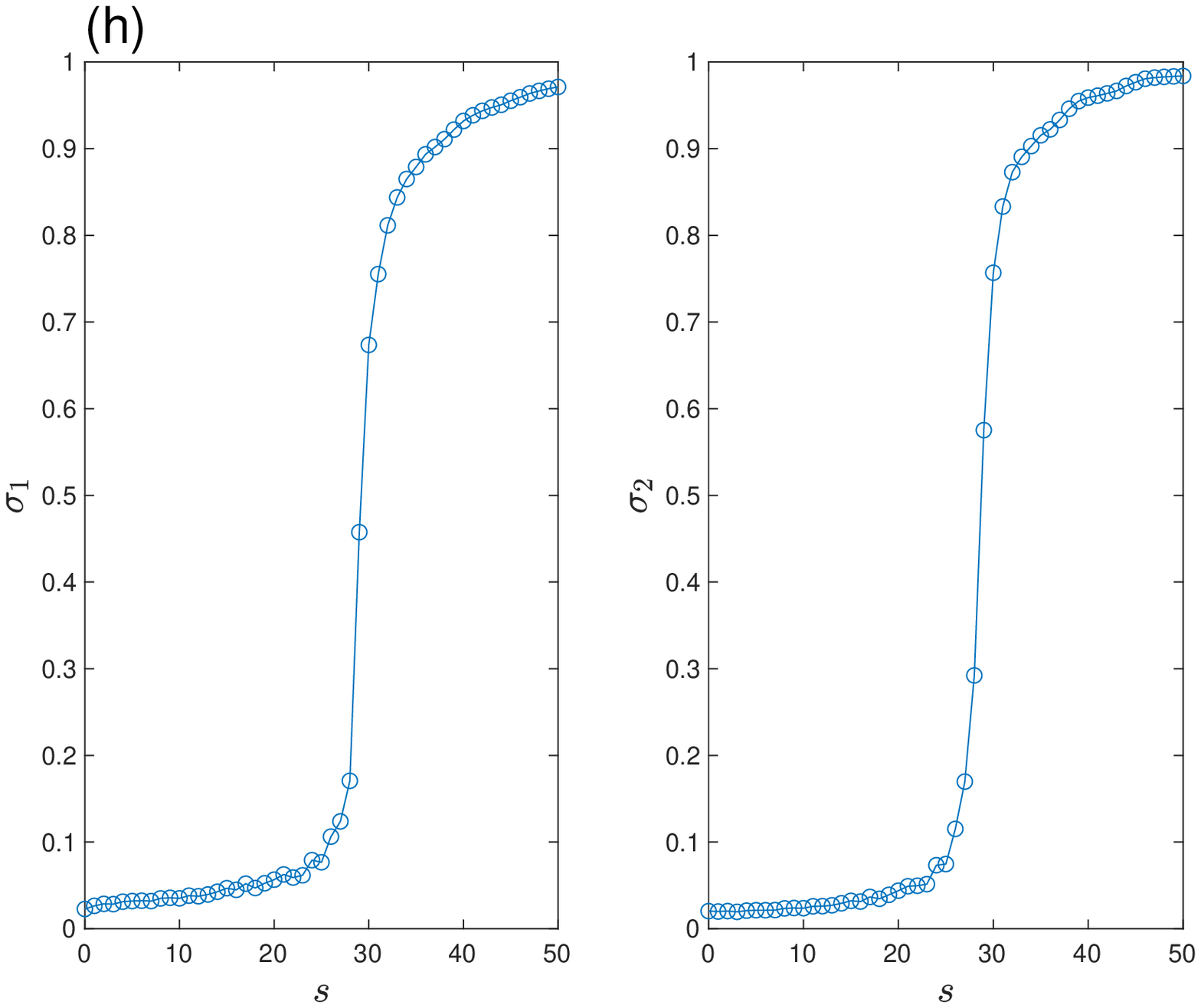} & \includegraphics[width=2.6cm]{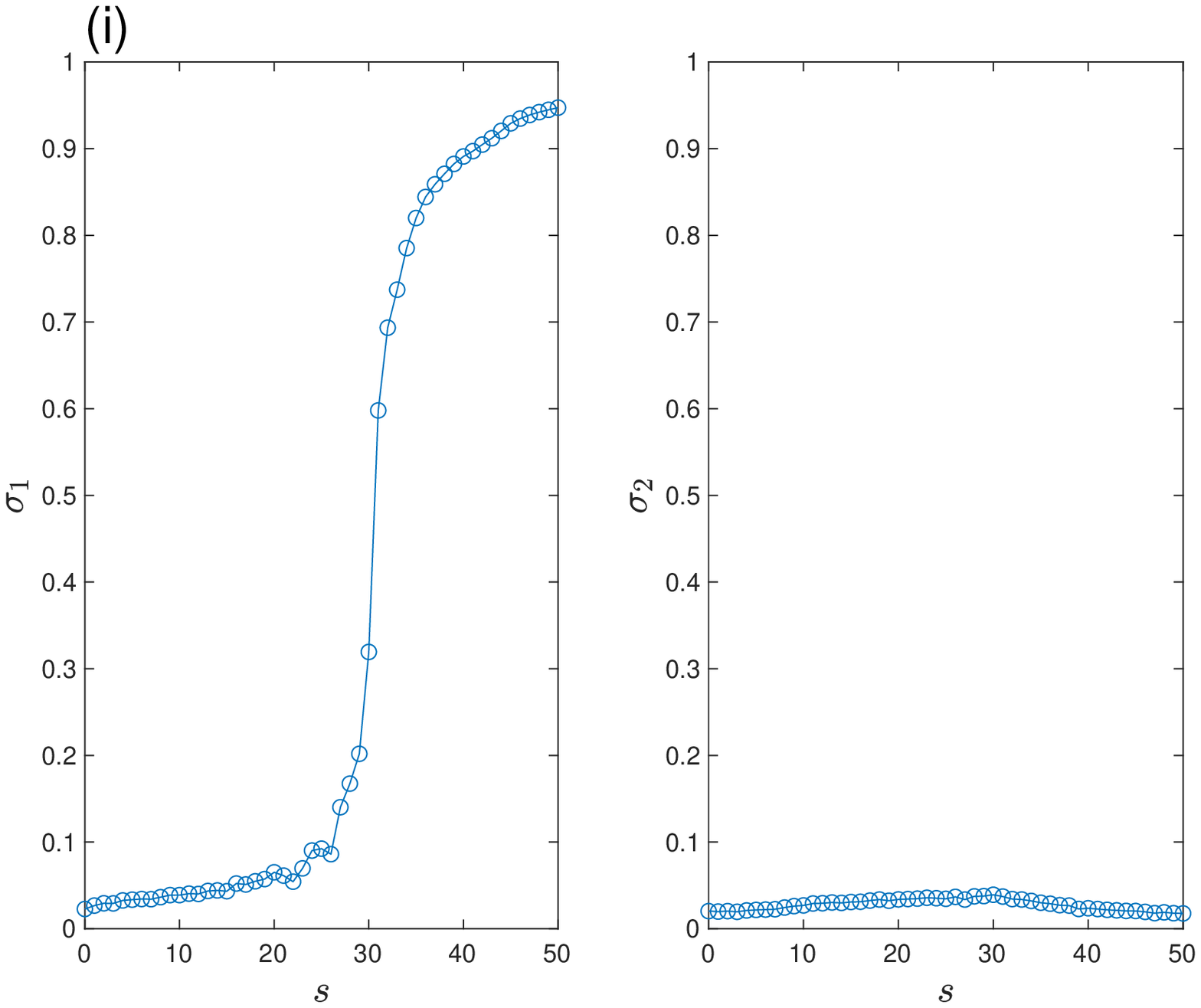}\tabularnewline
\includegraphics[width=2.6cm]{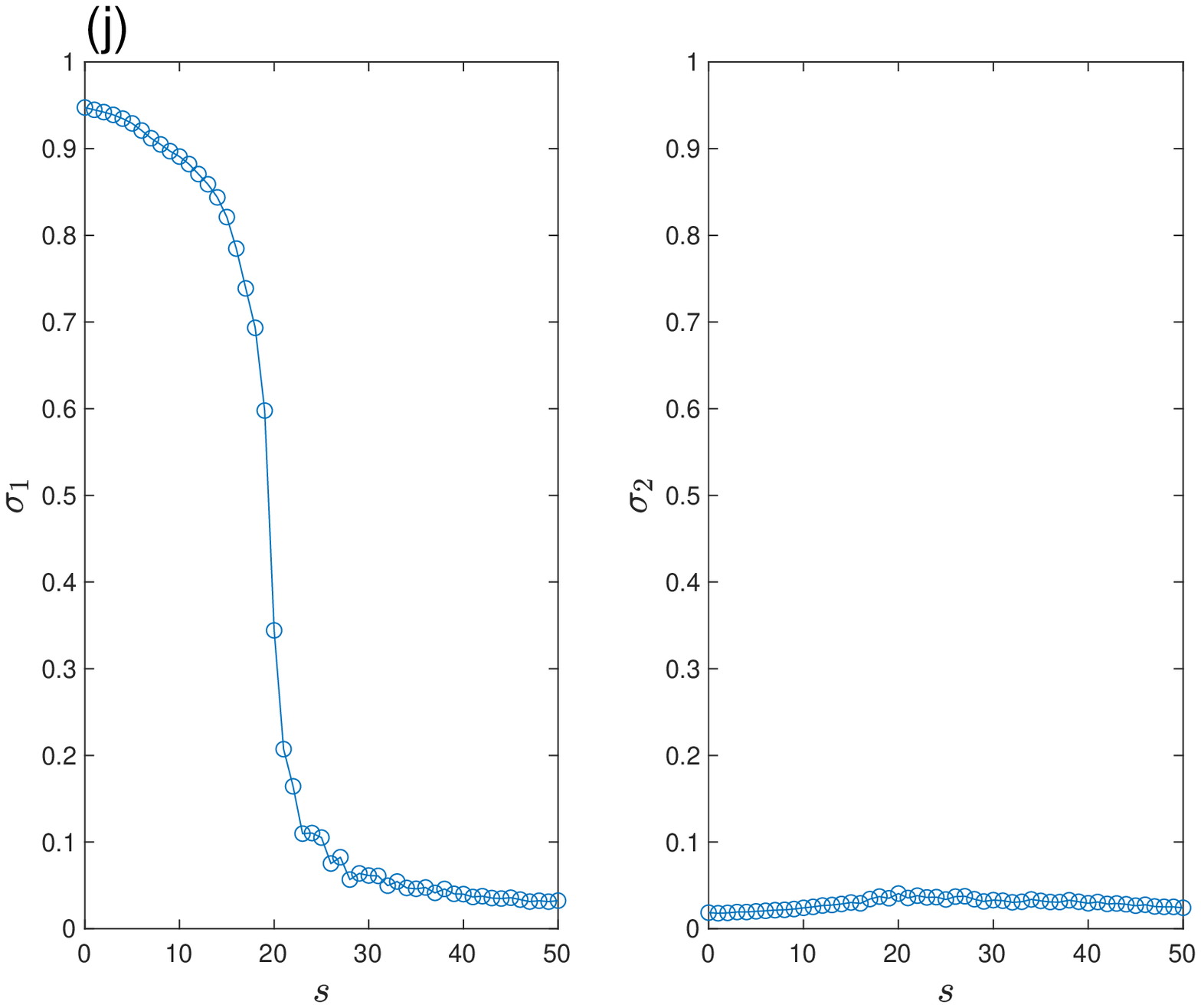} & \includegraphics[width=2.6cm]{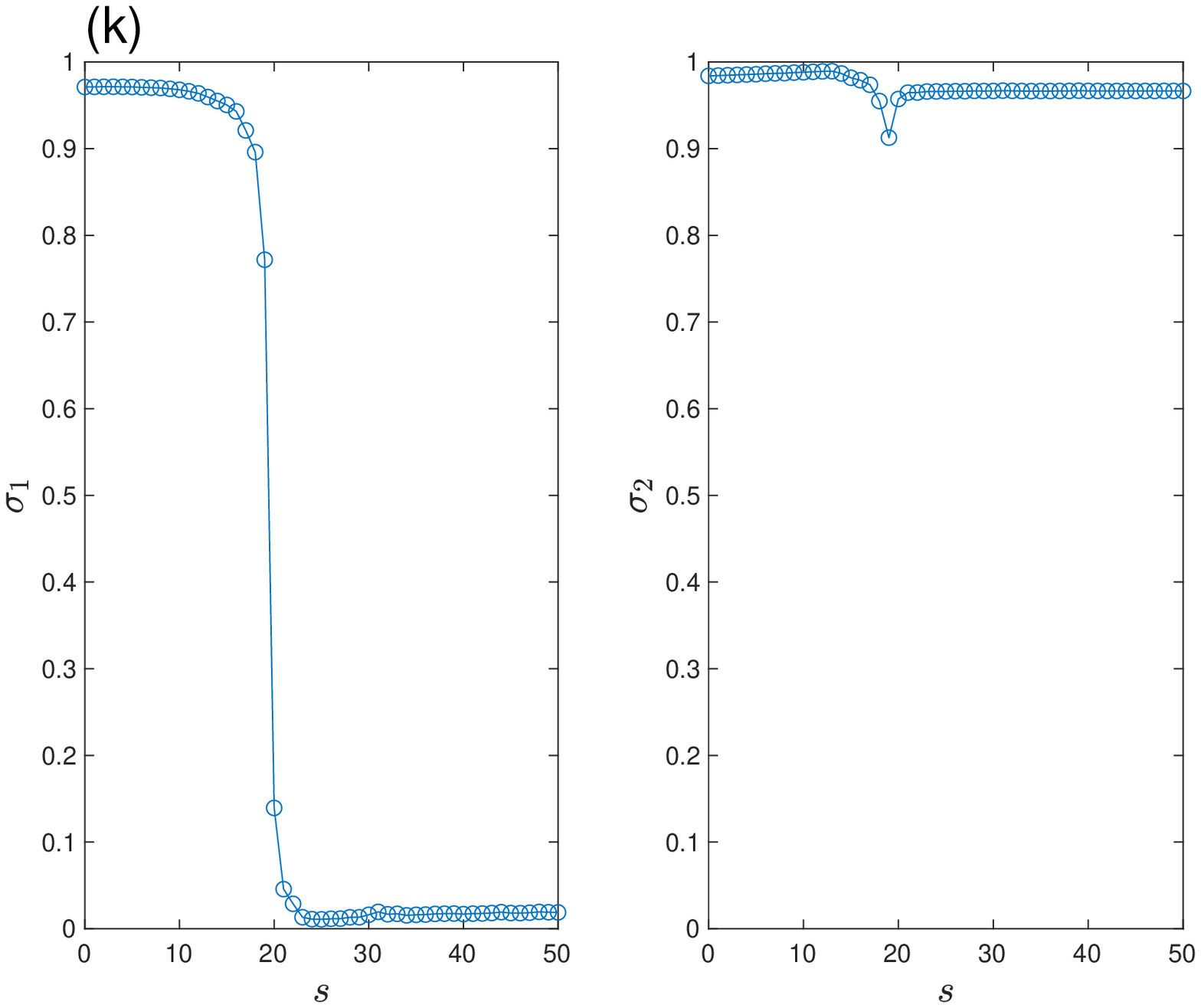} & \includegraphics[width=2.6cm]{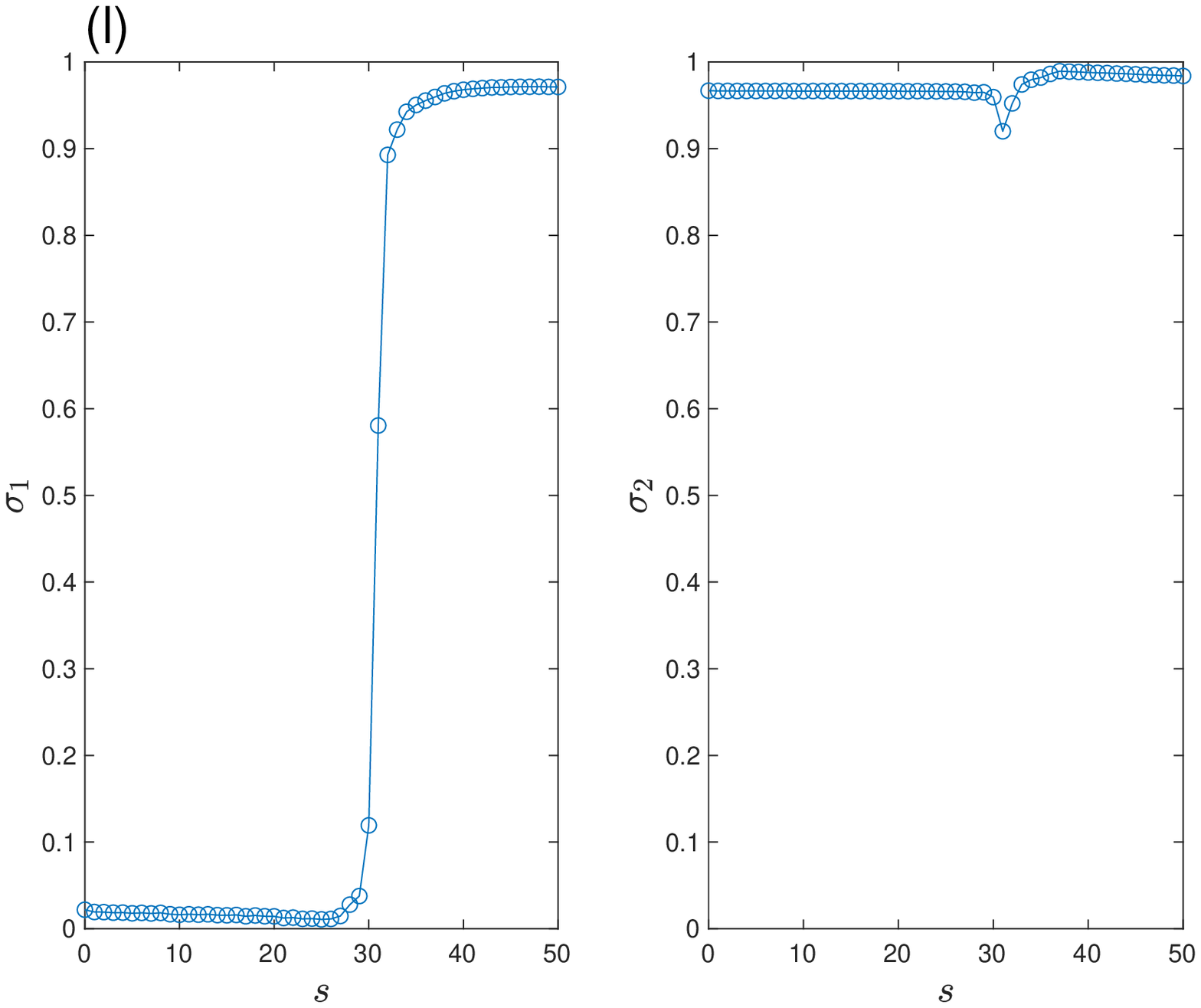}\tabularnewline
\end{tabular}

\caption{\emph{Switching diagram of $\sigma_{1,2}$.} The diagram on the top
illustrates the combination of $\sigma_{1}$ and $\sigma_{2}$ and
their interchanging process indexed as (a) to (l). Figures below draw
separately the transitions of $\sigma_{1}$ and $\sigma_{2}$ in the
aforementioned (a) to (l) with respect to step $s$ as in eqn. (\ref{eq:3}).
Refer to the text for the configuration of the experiment.}

\label{fig:FIG4}
\end{figure}

For the experiment, we demonstrate the switchings on $N=100$ oscillators
whose natural frequencies $\omega_{i1}=\omega_{i2}=\omega_{i}$ are
drawn from the standard Lorentzian distribution. Specifically, the
representative matrices are $M_{1}=\begin{bmatrix}-16 & -10\\
-10 & -20
\end{bmatrix}$ for mode $00$, $M_{2}=\begin{bmatrix}16 & 30\\
30 & 20
\end{bmatrix}$ for mode $01$, $M_{3}=\begin{bmatrix}16 & 30\\
30 & -4
\end{bmatrix}$ for mode $10$, $M_{4}=\begin{bmatrix}16 & 10\\
10 & 20
\end{bmatrix}$ for mode $11$, that will be morphing into one another during the
switching process. For each switching in (a)$\sim$(l) in FIG. \ref{fig:FIG4}.,
we simulate under dynamics (\ref{eq:1}) from $\theta_{i}(0)={\bf 0}$
and $A=M_{i}$ and break down its transition to $A=M_{j}$ into $51$
steps, $i,j\in\{1,2,3,4\}$; for each step, an average of the steady
state order parameters over time is evaluated as a data point on figures
(a)$\sim$(l), and the overall results of $\sigma_{1,2}$ are displayed
in FIG. \ref{fig:FIG4}. For switchings (a) to (l) other than (d),
we use linear interpolation in the $51$ steps to compensate the difference
between the matrices, i.e., 
\begin{equation}
M^{s+1}=M^{s}+\Lambda\label{eq:3}
\end{equation}
where $\Lambda=(M_{j}-M_{i})/50$ is the incremental matrix so that
$M^{0}=M_{i}=\begin{bmatrix}a_{11}^{i} & a_{12}^{i}\\
a_{21}^{i} & a_{22}^{i}
\end{bmatrix},M^{50}=M_{j}=\begin{bmatrix}a_{11}^{j} & a_{12}^{j}\\
a_{21}^{j} & a_{22}^{j}
\end{bmatrix}$. However for (d) which is from $10$ to $01$, the switching does
not happen with direct interpolation between $M_{3}=\begin{bmatrix}16 & 30\\
30 & -4
\end{bmatrix}$ and $M_{2}=\begin{bmatrix}16 & 30\\
30 & 20
\end{bmatrix}$; actually, one needs to extend the increasing of $a_{22}$ till around
$31$ to induce the switching. Therefore for (d), we first evenly
decrease $a_{11}$ while increasing $a_{22}$ of $M_{3}$ to $\begin{bmatrix}0 & 30\\
30 & 8
\end{bmatrix}$ in $26$ steps, then increase evenly both $a_{11}$ and $a_{22}$
in the remaining steps to $M_{2}$ which has allowed the desired switching
from $10$ to $01$.
 FIG. \ref{fig:FIG4}. verifies that, as the representative matrix
$M_{i}$ morphs into $M_{j}$ in the twelve scenarios, the system
completes a switching from $M_{i}$'s corresponding mode to that of
$M_{j}$'s with either continuous transitions or explosive transitions
to synchronization/desynchronization. Although we have not touched
on the problem of initial conditions in this work, consecutive switchings
with different combinations of (a)$\sim$(l) are produced as we continue
to apply interpolation, and obtain a sequence of $\sigma_{1,2}$ that
are ultimately varying with time, which might be a suggestion that
the alternating synchronization and desynchronization in the different
dimensions of the oscillators display capacity in coding information
in a binary fashion under specific choices of the coupling matrix.

To shed light on the phenomenon observed, we employ a self-consistency
analysis where the steady state assumption guarantees that $\dot{\Psi}_{1}=\Omega_{1},\dot{\Psi}_{2}=\Omega_{2}$,
with $\Omega_{1,2}$ being the symmetry centers of $g_{1}(\omega_{1})$
and $g_{2}(\omega_{2})$. Thus by introducing the rotating frame $\psi_{i1}=\theta_{i1}-\Omega_{1}t,\psi_{i2}=\theta_{i2}-\Omega_{2}t$,
eqn. (\ref{eq:2}) now turns into 
\begin{equation}
\begin{array}{c}
\dot{\psi}_{i1}=(\omega_{i1}-\Omega_{1})-(a_{11}\sigma_{1}\sin\psi_{i1}+a_{12}\sigma_{2}\sin\psi_{i2}),\\
\dot{\psi}_{i2}=(\omega_{i2}-\Omega_{2})-(a_{21}\sigma_{1}\sin\psi_{i1}+a_{22}\sigma_{2}\sin\psi_{i2}).
\end{array}\label{eq:4}
\end{equation}
Eqn. (\ref{eq:4}) suggests that when either of the order parameters
$\sigma_{1,2}$ is at the proximity of zero, e.g., when $\sigma_{2}\approx0$,
the other dimension of the oscillators get decoupled from its influence
and reduces to the dynamic of the classic Kuramoto model, that is
in our example, $\dot{\psi}_{i1}=(\omega_{i1}-\Omega_{1})-a_{11}\sigma_{1}\sin\psi_{i1}$,
and $\dot{\psi}_{i2}$ being driven by the external field $\sigma_{1}\sin\psi_{i1}$.
Another simplified scenario is when the system is devoid of desynchronized
oscillators on both dimensions as in the case (f) of FIG. \ref{fig:FIG4}.,
where $A$ is invertible and $\sigma_{1,2}\approx1$ at the beginning
of the transition. Denote $A^{-1}=\begin{bmatrix}b_{11} & b_{12}\\
b_{21} & b_{22}
\end{bmatrix}$; for $\psi_{i1}$ and $\psi_{i2}$ to have their respective attractors
at the same time, the synchronization domains are 

\noindent 
\begin{equation}
\begin{array}{c}
D_{1}=\left\{ (\omega_{i1},\omega_{i2}):\left|\frac{b_{11}(\omega_{i1}-\Omega_{1})+b_{12}(\omega_{i2}-\Omega_{2})}{\sigma_{1}}\right|<1\right\} ,\\
D_{2}=\left\{ (\omega_{i1},\omega_{i2}):\left|\frac{b_{21}(\omega_{i1}-\Omega_{1})+b_{22}(\omega_{i2}-\Omega_{2})}{\sigma_{2}}\right|<1\right\} .
\end{array}\label{eq:5}
\end{equation}
Solve for $\dot{\psi}_{i1}=0$ and $\dot{\psi}_{i2}=0$, the fixed
point for vector $\begin{bmatrix}\psi_{i1} & \psi_{i2}\end{bmatrix}$
is $\psi_{i1}^{*}=\arcsin(\frac{b_{11}(\omega_{i1}-\Omega_{1})+b_{12}(\omega_{i2}-\Omega_{2})}{\sigma_{1}}),\psi_{i2}^{*}=\arcsin(\frac{b_{21}(\omega_{i1}-\Omega_{1})+b_{22}(\omega_{i2}-\Omega_{2})}{\sigma_{2}})$
for $D_{1}\bigcap D_{2}\neq\emptyset$; under the premise that $\omega_{i1}=\omega_{i2}$,
these expressions can be further simplified with $\Omega_{1}=\Omega_{2}=\Omega$.
Since both $\dot{\psi}_{i1}$ and $\dot{\psi}_{i2}$ are synchronized
before the abrupt desynchronization, we calculate only synchronized
contribution to the order parameters $\sigma_{1,2}$. Apply the coordinate
transformation $\sin\Delta_{1}=\frac{a_{22}-a_{12}}{|A|\sigma_{1}}(\omega_{1}-\Omega),\sin\Delta_{2}=\frac{a_{11}-a_{21}}{|A|\sigma_{2}}(\omega_{2}-\Omega)$,
we then need to decide the new synchronization domain $D_{1}\cap D_{2}=\{(\Delta_{1},\Delta_{2}):-\eta_{1}<\Delta_{1}<\eta_{1},-\eta_{2}<\sin\Delta_{2}<\eta_{2}\}$
under this coordinate for $\omega_{1}=\omega_{2}$. In the continuum
limit $N\rightarrow\infty$, the order parameters are obtained as
\begin{align}
\sigma_{1}\{1-\left|\frac{|A|}{a_{22}-a_{12}}\right|\int_{-\eta_{1}}^{\eta_{1}}\cos^{2}\Delta_{1}g(\Omega+\frac{|A|\sigma_{1}}{a_{22}-a_{12}}\sin\Delta_{1})\nonumber \\
\cdot d\Delta_{1}\}=0,\nonumber \\
\\
\sigma_{2}\{1-\left|\frac{|A|}{a_{11}-a_{21}}\right|\int_{-\eta_{2}}^{\eta_{2}}\cos^{2}\Delta_{2}g(\Omega+\frac{|A|\sigma_{2}}{a_{11}-a_{21}}\sin\Delta_{2})\nonumber \\
\cdot d\Delta_{2}\}=0,\nonumber 
\end{align}
for which when $\left|\frac{\sigma_{1}}{a_{22}-a_{12}}\right|<\left|\frac{\sigma_{2}}{a_{11}-a_{21}}\right|$
, there is $\eta_{1}=\frac{\pi}{2},\eta_{2}=\arcsin(\left|\frac{a_{11}-a_{21}}{a_{22}-a_{12}}\right|\frac{\sigma_{1}}{\sigma_{2}})$;
alternatively when $\left|\frac{\sigma_{1}}{a_{22}-a_{12}}\right|>\left|\frac{\sigma_{2}}{a_{11}-a_{21}}\right|$,
there is $\eta_{1}=(\arcsin(\left|\frac{a_{11}-a_{21}}{a_{22}-a_{12}}\right|\frac{\sigma_{1}}{\sigma_{2}}))^{-1},\eta_{2}=\frac{\pi}{2}$.
Note that the integrated function is always positive, thus normally,
apart from the uniformly distributed solution $\sigma_{1}=\sigma_{2}=0$,
there exists another steady state solution suppose $\left|\frac{|A|}{a_{22}-a_{12}}\right|\neq0,$$\left|\frac{|A|}{a_{11}-a_{21}}\right|\neq0$.
This explains the abrupt transition to the desynchronized state for
$\sigma_{2}$, since when $\left|\frac{|A|}{a_{11}-a_{21}}\right|$
is gradually tuned to zero, the partially synchronized solution for
$\sigma_{2}$ vanishes and momentarily leaves $\sigma_{2}=0$ as the
only solution for the system. Actually, from the perspective of the
synchronization domains, the explanation covers a wider range of phenomena
in FIG. \ref{eq:4}. for which with $\omega_{1}=\omega_{2}=\omega$,
the integration domain is equivalently $D_{1}\bigcap D_{2},D_{1}:|\omega-\Omega|<\frac{\sigma_{1}\left||A|\right|}{\left|a_{22}-a_{12}\right|},D_{2}:|\omega-\Omega|<\frac{\sigma_{2}\left||A|\right|}{\left|a_{11}-a_{21}\right|}$,
that is, $D_{1}\bigcap D_{2}=\{\omega:|\omega-\Omega|<\min\{\left|\frac{|A|}{a_{22}-a_{12}}\right|,\left|\frac{|A|}{a_{11}-a_{21}}\right|\}\}.$
For (f), $\frac{|A|}{a_{11}-a_{21}}$ passes from being positive to
being negative one step earlier than $\frac{|A|}{a_{22}-a_{12}}$
and the right side of the expression $D_{2}$ approaches closely to
zero, which means almost all the oscillators other than those with
natural frequencies $\omega_{i1}=\omega_{i2}=\Omega$ are not entrained
by the mean-field of the second dimension, thus the order parameter
$\sigma_{2}$ has to see a significant drop. Indeed, for switchings
(a)$\sim$(l) other than (c), we have observed for $D_{1}$ and $D_{2}$
to cross zero almost simultaneously, and the narrower domain eventually
dominates the transition to synchronization, i.e., $\sigma_{1}\rightarrow1$
if $D_{1}\subset D_{2}$ at the end of the transition. 

In this Letter, we present a novel coupling mechanism with the $2\times2$
real matrices on the two-dimensional Kuramoto oscillators and uncover
distinct phenomena it induces with different configurations and variations
of the coupling matrix. One thing we have discovered is that, the
positivity or negativity of the eigenvalues of the matrix indicates
the tendency of the two dimensions of the oscillators to set into
synchronization or desynchronization, as the four elements of the
matrix are scaled by a positive factor sufficiently large. Since the
synchronization and desynchronization of the two dimensions are separated,
their combinations suggest four qualitatively distinct modes of the
system that are possible to switch between one and another through
the variations of the coupling matrix. The switching between synchronization
and desynchronization in the two dimensions with respect to time displays
potential in information coding and memory storage \citep{fell2011role}
and imitates other phenomenology in biology like the unihemispheric
slow-wave sleep of dolphins \citep{mukhametov1977interhemispheric},
where the two hemispheres of the dolphin alternate between resting
and waking during its sleep, the two behaviors existing independently
at the same time. Apart from the experiments with $2\times2$ real
matrices presented in this work, we have also simulated on $3\times3$
real matrices with various eigenvalue arrangements and recovered the
potential eight combinations of the order parameters $\sigma_{1,2,3}$,
suggesting the proposed model to be quite generalizable into even
higher dimensions.

\bibliographystyle{apsrev4-1}
\bibliography{lib_K2dim}

\end{document}